\documentclass{article}
\usepackage{sylvie}

\title{A New Technique for Reachability of States in Concatenation Automata}
\author{Sylvie Davies}
\date{\small University of Waterloo\\Department of Pure Mathematics\\{\tt sldavies@uwaterloo.ca}}

\newcommand{\stc}{\operatorname{sc}}
\newcommand{\psupp}{\operatorname{psupp}}
\newcommand{\dist}{\sim}

\newcommand{\tr}[1]{\overset{#1}{\longrightarrow}}

\begin{document}
\maketitle

\begin{abstract}
We present a new technique for demonstrating the reachability of states in deterministic finite automata representing the concatenation of two languages.
Such demonstrations are a necessary step in establishing the state complexity of the concatenation of two languages, and thus in establishing the state complexity of concatenation as an operation.
Typically, ad-hoc induction arguments are used to show particular states are reachable in concatenation automata.
We prove some results that seem to capture the essence of many of these induction arguments.
Using these results, reachability proofs in concatenation automata can often be done more simply and without using induction directly.
\end{abstract}

\section{Introduction}
\label{sec:intro}
Formal definitions are postponed until Section \ref{sec:pre}.

The \e{state complexity} of a regular language $L$, denoted $\stc(L)$, is the least number of states needed to recognize the language with a deterministic finite automaton. The \e{state complexity of an operation} on regular languages is the worst-case state complexity of the result of the operation, expressed as a function of the maximal allowed state complexity of the input languages. For example, suppose $L$ is a language of state complexity at most $m$, and $K$ is a language of state complexity at most $n$. It is known that the intersection $L \cap K$ has state complexity at most $mn$, and that this upper bound can be attained. Thus, we say that the \e{state complexity of intersection} is the function $(m,n) \mapsto mn$.

To establish the state complexity of an operation, there are two steps.
First, one derives an upper bound. For example, in the case of intersection, if the input languages $L$ and $K$ have state complexity at most $m$ and at most $n$ respectively, then the standard direct product construction gives an automaton for $L \cap K$ with $mn$ states, leading to the aforementioned upper bound of $mn$.
Next, one searches for witnesses to the upper bound, that is, languages which attain the upper bound for each value of $m$ and $n$. In the case of intersection, this means for each pair $(m,n)$, one must find a pair of languages $(L_m,K_n)$ with $\stc(L_m) \le m$ and $\stc(K_n) \le n$ such that $\stc(L_m \cap K_n) = mn$.

One must not only find these witnesses but also \e{prove} that the desired state complexity bound is reached.
Such proofs are the subject of this paper. We are interested in the case where the operation is \e{concatenation} of languages. 
We assume that one is working within some subclass of the regular languages, and has derived an upper bound $f(m,n)$ for the worst-case state complexity of concatenation within this subclasses.
We also assume one has found (by computer search or some other means) candidate witnesses for this upper bound, in the form of two sequences of languages $(L_m : m \ge 1)$ and $(K_n : n \ge 1)$ such that $\stc(L_m) \le m$ and $\stc(K_n) \le n$. 
The goal is to prove that for each pair $(m,n)$, the concatenation $L_mK_n$ has state complexity $f(m,n)$.
We may divide such a proof into three steps:
\be
\item
Construct an automaton $\cA$ for $L_mK_n$ in the standard way.
\item
Show that $\cA$ contains at least $f(m,n)$ reachable states.
\item
Show exactly $f(m,n)$ reachable states in $\cA$ are pairwise distinguishable.
\ee
We present a new technique for dealing with step (2) of this process. The standard way to construct a deterministic finite automaton $\cA$ for the concatenation of two languages yields an automaton in which the states are \e{sets}; to show a particular set is reachable, one typically proceeds by induction on the size of the set.
We prove a result that seems to generalize many of these ad-hoc induction arguments, and can be used to establish reachability of sets without directly using induction.
Additionally, we prove some helpful lemmas that make our main result easier to apply.

We demonstrate our technique by applying it to a variety of concatenation witnesses taken from the literature. The state complexity of concatenation has been studied in the class of all regular languages, as well as many subclasses. Table \ref{tb:results} lists some examples of subclasses that have been studied, and the state complexity of concatenation in each subclass. 
See the cited papers for definitions of each subclass and derivations/proofs of each complexity.
The complexities listed are ``restricted'' complexities, that is, they are computed under the assumption that both inputs to the concatenation operation share the same alphabet. ``Unrestricted'' state complexity of concatenation (where the inputs may be languages over different alphabets) has also been studied, and will be discussed later, but is not included in the table.


\begin{table}[h]
\centering\scriptsize
\begin{tabular}{|c|c||c|c|}
\hline
Subclass                          & Complexity            &Subclass        & Complexity                    \\
\hline
Regular~\cite{Brz13,BrSi17,Mas70,YZS94} & $\mathbf{(m-1)2^n + 2^{n-1}}$ 
&
Prefix-closed~\cite{BJZ14,BrSi17} &$\mathbf{(m+1)2^{n-2}}$ \\
\hline
Unary~\cite{Nic99,PiSh02,YZS94} & ${\sim}mn$ (asymptotically)
&
Prefix-free~\cite{BrSi17,HSW09,JiKr10} & $m+n-2$ \\
\hline
Finite unary~\cite{CCSY01,Yu01} & $m+n-2$
&
Suffix-closed~\cite{BJZ14,BrSi17a}     & $mn-n+1$ \\
\hline
Finite binary~\cite{CCSY01}              & $\mathbf{(m-n+3)2^{n-2}-1}$   
&
Suffix-free~\cite{BrSi17a,HaSa09} & $\mathbf{(m-1)2^{n-2}+1}$ \\
\hline
Star-free~\cite{BrLi12}           & $\mathbf{(m-1)2^n + 2^{n-1}}$ 
&
Right ideal~\cite{BDL16,BJL13,BrSi17} & $\mathbf{m+2^{n-2}}$   \\
\hline
Non-returning~\cite{BrDa17,EHJ16}      & $\mathbf{(m-1)2^{n-1}+1}$
&
Left ideal~\cite{BDL16,BJL13,BrSi17a} & $m+n-1$         \\
\hline
\end{tabular}
\caption{Subclasses of regular languages and the state complexity of the concatenation operation within each subclass. {\bf Bold} type indicates that the complexity grows exponentially in terms of $n$.}
\label{tb:results}
\end{table}
If the state complexity of concatenation grows exponentially with $n$ (indicated in Table \ref{tb:results} by {\bf bold} type), it is typical to use an induction argument to prove the desired number of states is reachable. It is cases like this in which our technique is most likely to be useful.
We selected 16 concatenation witnesses, all from subclasses in which the state complexity of concatenation is exponential in $n$, and tried to apply our technique to these witnesses. In many cases we were able to produce shorter and simpler proofs than the original authors, and we only found two cases in which our technique did not work or was not useful.
This suggests that our technique is widely applicable and should be considered as an viable alternative to the traditional induction argument when attempting reachability proofs in concatenation automata.

The rest of the paper is structured as follows. 
Section \ref{sec:pre} contains background material and definitions needed to understand the paper.
Section \ref{sec:main} describes our new technique and proves the relevant results.
Section \ref{sec:ex} contains examples of our technique applied to numerous concatenation witnesses from the literature.
Section \ref{sec:con} concludes the paper.

\section{Preliminaries}
\label{sec:pre}
\subsection{Relations and Functions}
A \e{binary relation} $\rho$ between $X$ and $Y$ is a subset of $X \times Y$. If $\rho \subseteq X \times Y$ and $\tau \subseteq Y \times Z$, the \e{composition} of $\rho$ and $\tau$ is the relation 
\[ \rho\tau = \{(x,z) \in X \times Z : \text{ there exists } y \in Y \text{ such that } (x,y) \in \rho \text{ and } (y,z) \in \tau\}. \]
For $x \in X$ and $\rho \subseteq X \times Y$, the \e{image} of $x$ under $\rho$ is the set $x\rho = \{ y \in Y : (x,y) \in \rho\}$.
For $x \not\in X$ we define $x\rho = \emp$.
The \e{converse} of a binary relation $\rho \subseteq X \times Y$ is the relation $\rho^{-1} = \{ (y,x) : (x,y) \in \rho\} \subseteq Y \times X$.
The set $y\rho^{-1} = \{ x \in X : (x,y) \in \rho\}$ is called the \e{preimage} of $y$ under $\rho$. Elements of this set are callled \e{preimages} of $y$; for example, if $x \in y\rho^{-1}$ we say that $x$ is a preimage of $y$.

If we write $\cP(S)$ for the power set of a set $S$ (that is, the set of all subsets of $S$), then we can view $\rho$ as a map $\rho \co X \ra \cP(Y)$.
We may also \e{extend $\rho$ by union} to a map $\rho \co \cP(X) \ra \cP(Y)$ as follows: for $S \subseteq X$, we define
\[ S\rho = \bigcup_{s \in S} s\rho.  \]
We thus have two ways to make sense of an expression like $x\rho\tau$: it is the image of $x$ under the composite relation $\rho\tau \subseteq X \times Z$, and it is also the image of the set $x\rho \subseteq Y$ under the map $\tau \co \cP(Y) \ra \cP(Z)$. Additionally, we have a way to make sense of a composition $\rho\tau \co X \ra \cP(Z)$ of maps $\rho \co X \ra \cP(Y)$ and $\tau \co Y \ra \cP(Z)$: take the composition of the corresponding relations.

A \e{function} $f \co X \ra Y$ is a binary relation $f \subseteq X \times Y$ such that $|xf| = 1$ for all $x \in X$. Following our notation for binary relations, we write functions to the \e{right} of their arguments.
Composition of functions is defined by composing the corresponding relations. Thus the order of composition is \e{left-to-right}; in a composition $fg$, first $f$ is applied and then $g$.

A \e{transformation} of a set $X$ is a function $t \co X \ra X$, that is, a function from $X$ into itself.
We say $t$ is a \e{permutation} of $X$ if $Xt = X$. We say $t$ \e{acts as a permutation} on $S \subseteq X$ if $St = S$.
If $t$ acts as a permutation on $S$, then every element of $S$ has at least one preimage under $t$, that is, for all $s \in S$, the set $st^{-1} = \{ x \in X : xt = s\}$ is non-empty. 

A \e{cyclic permutation} of a set $\{x_1,\dotsc,x_k\} \subseteq X$ is a permutation $p$ such that $x_ip = x_{i+1}$ for $1 \le i < k$, $x_kp = x_1$, and $xp = x$ for all $x \in X \setminus \{x_1,\dotsc,x_k\}$. We denote such a permutation as $(x_1,\dotsc,x_k)$. A cyclic permutation of a two-element set is called a \e{transposition}. The identity transformation is denoted $\id$.

The notation $(S \ra x)$ for $S \subseteq X$ and $x \in X$ denotes a transformation that sends every element of $S$ to $x$ and fixes every element of $S \setminus X$.
For example, $(\{i\} \ra j)$ denotes a transformation that maps $i$ to $j$ and fixes everything else.
The transformation $(X \ra x)$ is a constant transformation that maps every element of $X$ to $x$.

In the case where $X = \{1,2,\dotsc,n\}$, the notation $({}_i^j x \ra x+1)$ denotes a transformation such that for each $x$ with $i \le x \le j$, the transformation sends $x$ to $x+1$, and every other $x$ is fixed.
For example, the transformation $({}_2^{n-1} x \ra x+1)$ fixes 1, sends $x$ to $x+1$ for $2 \le x \le n-1$, and fixes $n$.
The notation $({}_i^j x \ra x-1)$ is defined similarly.


\subsection{Automata}
\label{sec:def:aut}
A \e{finite automaton} (FA) is a tuple $\cA = (Q,\Sig,T,I,F)$ where $Q$ is a finite set of \e{states}, $\Sig$ is a finite set of \e{letters} called an \e{alphabet}, $T \subseteq Q \times \Sig \times Q$ is a set of \e{transitions}, $I \subseteq Q$ is a set of \e{initial states}, and $F \subseteq Q$ is a set of \e{final states}.

We now define a binary relation $T_w \subseteq Q \times Q$ for each $w \in \Sig^*$.
Define $T_\eps = \{(q,q) : q \in Q\}$; in terms of maps, this is the identity map on $Q$.
For $a \in \Sig$, define $T_a = \{ (p,q) \in Q \times Q : (p,a,q) \in T\}$.
For $w = a_1\dotsb a_k$ with $a_1,\dotsc,a_k \in \Sig$, define $T_w = T_{a_1} \dotsb T_{a_k}$.
The relation $T_w$ is called the \e{relation induced by $w$}.
The set $\{T_w : w \in \Sig^*\}$ is a monoid under composition, called the \e{transition monoid} of $\cA$.
For technical reasons, if $w$ is a word but is \e{not} a word over $\Sig$, we define $T_w$ to be the empty relation: $qT_w = \emp$ for all $q \in Q$.

Observe that the set of transitions of an FA is determined by the relations $T_a$. Furthermore, each relation $T_a$ is determined by the set of images $qT_a$, where $q \in Q$. Hence we often define the transitions of an FA by specifying the images $qT_a$ for each $q \in Q$ and $a \in \Sig$.

If $\cA = (Q,\Sig,T,I,F)$ is a finite automaton such that $|I| = 1$ and $T_a$ is a \e{function} for each $a \in \Sig$,  we say $\cA$ is \e{deterministic}. We abbreviate ``deterministic finite automaton'' to DFA. 

Let $\cA = (Q,\Sig,T,I,F)$ be an FA.
A word $w \in \Sig^*$ is \e{accepted} by $\cA$ if we have $IT_w \cap F \ne \emp$. If $\cA$ is a DFA with $I = \{i\}$, this condition becomes $iT_w \in F$. The \e{language} of $\cA$, denoted $L(\cA)$, is the set of all words it accepts. 
Languages of FAs are called \e{regular languages}. 
A sequence of transitions $(q_0,a_1,q_1),(q_1,a_2,q_2)\dotsc,(q_{k-1},a_k,q_k)$ with $w = a_1\dotsb a_k$ is called a \e{path} from $q_0$ to $q_k$ with label $w$, and the path is \e{accepting} if $q_0 \in I$ and $q_k \in F$.
The FA $\cA$ accepts a word $w$ if and only if there is an accepting path with label $w$.
We write $p \tr{w} q$ to mean that there is a path from $p$ to $q$ with label $w$.

Given two regular languages $L$ and $K$ with DFAs $\cA = (Q^\cA,\Sig^\cA,T^\cA,i^\cA,F^\cA)$ and $\cB = (Q^\cB,\Sig^\cB,T^\cB,i^\cB,F^\cB)$, we may construct an FA $\cA\cB = (Q,\Sig,T,I,F)$ that accepts the concatenation $LK$ as follows:
\bi
\item
$Q = Q^\cA \cup Q^\cB$. We assume without loss of generality that $Q^\cA \cap Q^\cB = \emp$.
\item
$\Sig = \Sig^\cA \cup \Sig^\cB$.
\item
$T = T^\cA \cup T^\cB \cup \{(q,a,i^\cB) : q T^\cA_a \in F^\cA, a \in \Sig^\cA \}$.
\item
$I = \{i^\cA\}$ if $i^\cA \not\in F^\cA$, and otherwise let $I = \{i^\cA,i^\cB\}$.
\item
$F = F^\cB$.
\ei
\bp
The FA $\cA\cB$ accepts the concatenation $LK$.
\ep
\bpf
Suppose $w \in LK$; we want to show that $w$ is accepted by $\cA\cB$.
We can write $w = uv$ with $u \in L$ and $v \in K$.
There are two cases: $u$ can be empty or non-empty.
If $u = \eps$ then $\eps \in L$, so $i^\cA \in F^\cA$.
Thus $I = \{i^\cA,i^\cB\}$.
It follows that $IT_v \supseteq i^\cB T_v \ni i^\cB T^\cB_v$, which is final since $v \in K$.
If $u \ne \eps$ we can write $u = xa$ for some word $x$ and letter $a$.
Then $IT_{xa} \supseteq i^\cA T^\cA_{xa}$, which contains an element of $F^\cA$ since $xa \in L$.
It follows that $IT_x \supseteq i^\cA T^\cA_x$ contains some state $q$ such that $qT^\cA_a \in F^\cA$.
Thus $IT_{xa} \supseteq qT_a \ni i^\cB$, and so $IT_{xav} = IT_w \ni i^\cB T^\cB_v$, which is final since $v \in K$. 

Conversely, let $w$ be accepted by $\cA\cB$.
Choose an accepting path for $w$. There are two cases: either this accepting path starts from $i^\cB$, or it contains exactly one transition leading from a state of $Q^\cA$ to $i^\cB$. 
In the first case we must have $I = \{i^\cA,i^\cB\}$, and so $i^\cA \in F^\cA$.
This means $\eps \in L$. 
If an accepting path starts from $i^\cB$, all transitions on the path must belong to $T^\cB$. Thus we have $i^\cB T^\cB_w \in F^\cB$. It follows that $w \in K$, and so $w \in LK$.
In the second case, where the path contains exactly one transition from $Q^\cA$ to $i^\cB$, note that this transition must be of the form $(q,a,i^\cB)$ where $qT^\cA_a \in F^\cA$. Note also that every transition before this one lies in $T^\cA$, and every transition after lies in $T^\cB$.
Write $w = uav$; then $i^\cA T^\cA_{ua} = qT^\cA_a \in F^\cA$, so $ua \in L$. 
Also, $i^\cB T^\cB_v$ must be final, or our path would not be accepting.
Thus $v \in K$, and $uav = w \in LK$.
\epf

We are interested in the deterministic state complexity of concatenation, so we convert the FA $\cA\cB = (Q,\Sig,T,I,F)$ to a DFA recognizing the same language. The DFA we use is $\cC = (\cP(Q),\Sig,T,I,F_0)$, where $S \subseteq Q$ is in $F_0$ if $S \cap F \ne \emp$. Since each relation $T_a$ can be viewed as a function from $\cP(Q)$ to itself, and there is a unique initial state $I \in \cP(Q)$, this automaton is indeed deterministic. Since $IT_w \in F_0$ if and only if $IT_w \cap F \ne \emp$, we see that $\cC$ recognizes the same language as $\cA\cB$. We call $\cC$ the \e{concatenation DFA} for $\cA$ and $\cB$.

We make some observations and introduce some conventions to make it easier to work with the concatenation DFA.
\bi
\item
Since we are assuming $\cA$ and $\cB$ are DFAs, the only reachable states in $\cC$ have the form $S^\cA \cup S^\cB$, where $S^\cA \subseteq Q^\cA$, $S^\cB \subseteq Q^\cB$, and $|S^\cA| \le 1$.
Without loss of generality, we can assume the state set of $\cC$ consists of states of this form, rather than all of $\cP(Q)$. 
\item
We mark the states of $\cA$ with primes so they can be distinguished from the states of $\cB$. So a variable named $p$ or $q$ generally means a element of $Q^\cB$, while $p'$ or $q'$ means an element of $Q^\cA$.
\item
We identify the set $S^\cA \cup S^\cB$ with the ordered pair $(S^\cA,S^\cB)$. Hence we can view the states of $\cC$ as these ordered pairs. Reachable states are either of the form $(\emp,S)$ or $(\{q'\},S)$ with $q' \in Q^\cA$, $S \subseteq Q^\cB$. 
\item
For convenience, we frequently make no distinction between singleton sets and the elements they contain, and so write $(q',S)$ rather than $(\{q'\},S)$.
\item
Rather than $T_w$, $T^\cA_w$ and $T^\cB_w$, we simply write $w$ when it is clear from context which relation is meant. For example, $(q',S)w$ means $(q',S)T_w$ since $(q',S)$ is a state of $\cC$, and thus $T_w$ is the natural relation to apply. From our convention for marking the states of $\cA$ and $\cB$ with primes, one can infer that $q'w$ means $q'T^\cA_w$ and $qw$ means $qT^\cB_w$.
\item
Rather than $i^\cA$ and $i^\cB$, let $1'$ denote the initial state of $\cA$ and let $1$ denote the initial state of $\cB$.
We also assume without loss of generality that $Q^\cA = \{1',2',\dotsc,m'\}$ and $Q^\cB = \{1,2,\dotsc,n\}$ for some $m$ and $n$.
\ei
Under these conventions, the transitions of $\cC$ can be described as follows:
\[
(q',S)a = \begin{cases}
(\emp,Sa), & \text{if $a \in \Sig^\cB \setminus \Sig^\cA$;} \\
(q'a,\emp), & \text{if $a \in \Sig^\cA \setminus \Sig^\cB$ and $q'a \not\in F^\cA$;} \\
(q'a,1), & \text{if $a \in \Sig^\cA \setminus \Sig^\cB$ and $q'a \in F^\cA$;} \\
(q'a,Sa),  & \text{if $a \in \Sig^\cA \cap \Sig^\cB$ and $q'a \not\in F^\cA$;} \\
(q'a,Sa \cup 1),  & \text{if $a \in \Sig^\cA \cap \Sig^\cB$ and $q'a \in F^\cA$.} 
\end{cases}
\]
Recall that $T^\cA_w$ is the empty relation if $w$ is not a word over $\Sig^\cA$, and similarly for $\cB$. Thus the transitions admit a simpler description:
\[
(q',S)a = \begin{cases}
(q'a,Sa \cup 1),  & \text{if $a \in \Sig^\cA$ and $q'a \in F^\cA$;}\\
(q'a,Sa),  & \text{otherwise.} 
\end{cases}
\]

\subsection{State Complexity}
We say a DFA $\cA$ is \e{minimal} if it has the least number of states among all DFAs that recognize $L(\cA)$.
It is well known that each regular language has a unique minimal DFA (up to renamings of the states).
The \e{state complexity} of a regular language $L$, denoted $\stc(L)$, is the number of states in its minimal DFA.

There is a subtlety in this definition, arising from the fact that there are two common ways to define equality of functions.
One way is to say that functions are simply certain sets of ordered pairs, and are equal if they are equal as sets. 
The other way is to say that functions are triples $(f,D,C)$, where $D$ is the domain of the function and $C$ is the codomain, and thus two functions are equal if they are equal as sets \e{and} have the same domain and codomain.
Since words over alphabets are formally defined as functions, the first viewpoint implies that two words over distinct alphabets can be equal, while the second viewpoint implies two words over distinct alphabets are always distinct.
We call the first viewpoint the \e{unrestricted viewpoint}, and the second the \e{restricted viewpoint}, since the second viewpoint has more restrictive conditions for function equality.  

Now, consider how this affects the state complexity of the language $L = \{a\}^*$ over alphabet $\{a,b\}$. The smallest DFA with alphabet $\{a,b\}$ that recognizes $L$ has two states; a second state is necessary to exclude the words that contain $b$. Thus in the restricted viewpoint, the state complexity of $L$ is two. But in the unrestricted viewpoint, $L$ is equal to the language $\{a\}^*$ over alphabet $\{a\}$, which is recognized by a one-state DFA; thus the state complexity of $L$ is one.

The following characterization of minimality is useful.
Let $\cD = (Q,\Sig,T,i,F)$ be a DFA. 
A state $q \in Q$ is \e{reachable} if $iw = q$. 
For $p,q \in Q$, we say $q$ is \e{reachable from $p$} if $pw = q$.
Two states $p,q \in Q$ are \e{indistinguishable} if they are equivalent under the following equivalence relation: $p \dist q$ if for all $w \in \Sig^*$, we have $pw \in F \iff qw \in F$.
Otherwise they are \e{distinguishable} by some word $w$ such that $pw \in F \iff qw \not\in F$.
In the restricted viewpoint, a DFA is minimal if and only if all of its states are reachable and pairwise distinguishable.
In the unrestricted viewpoint, we also require that the DFA has an alphabet of minimal size.

Let $\circ$ be a binary operation on regular languages. The \e{state complexity of the operation $\circ$} is the following function, where $m$ and $n$ are positive integers:
\[ (m,n) \mapsto \max\{ \stc(L \circ K) : \stc(L) \le m, \stc(K) \le n\}. \]
This is the worst-case state complexity of the result of the operation, expressed as a function of the maximal allowed state complexities of the input languages. As with state complexity of languages, this definition differs depending on whether we adopt the restricted or unrestricted viewpoint, but the consequences are farther-reaching.

In the restricted viewpoint, we must assume that the inputs to the binary operation are languages over a \e{common alphabet}. The restricted viewpoint considers words over different alphabets to be always distinct, so it generally does not make sense to perform binary operations on languages over different alphabets. For example, if we take the language $L = \{ab\}$ over alphabet $\{a,b\}$, and the language $K = \{ab\}$ over alphabet $\{a,b,c\}$, the union $L \cup K$ contains two distinct elements both representing the word $ab$. This set $L \cup K$ is arguably not a language at all, since it cannot be written as a set of words over a single alphabet. Thus when computing the \e{restricted state complexity} of binary operations, we only consider inputs with the same alphabet. 

In the unrestricted viewpoint, there is no issue in allowing the input languages to have different alphabets. Thus when computing the \e{unrestricted state complexity} of binary operations, we consider all possible inputs to the operation, including pairs of languages with different alphabets. Allowing for different alphabets makes unrestricted state complexity slightly more complicated to compute. In fact, for many years, papers on operational state complexity only considered restricted state complexity. Unrestricted state complexity was first studied by Brzozowski~\cite{Brz16} in 2016, who pointed out that the restriction to common alphabets is artificial and can be removed.

Let us derive an upper bound for the restricted and unrestricted state complexities of the concatenation operation. We begin with two DFAs $\cA$ and $\cB$ that have $m$ and $n$ states respectively. The number of reachable states in the concatenation DFA $\cC$ for $\cA$ and $\cB$ gives an upper bound for the state complexity of $L(\cA)L(\cB)$.
Recall that reachable states have the form $(S^\cA,S^\cB)$, where $S^\cA \subseteq Q^\cA$, $S^\cB \subseteq Q^\cB$ and $|S^\cA| \le 1$. Since $|Q^\cA| = m$, there are $m+1$ possible values for $S^\cA$ (each of the singletons and the empty set).
Since $|Q^\cB| = n$, there are $2^n$ possible values for $S^\cB$.
However, if $S^\cA = \{f\}$ for a final state $f \in F^\cA$, the transition structure of $\cC$ tells us that we must have $1 \in S^\cB$. 
Thus if $|F^\cA| = k$, then there are at most $(m+1-k)2^n$ states with a non-final state or the empty set in the first component, and $k2^{n-1}$ states with a final state in the first component.
It follows there are at most $(m+1-k)2^n+k2^{n-1}$ reachable states in $\cC$.
This is maximized by taking $k = 1$, giving an upper bound of $m2^n+2^{n-1}$ in the unrestricted case.
For the restricted case, note that we cannot get the empty set in the first component, since this requires using a letter in $\Sig^\cB \setminus \Sig^\cA$.
Thus we get an upper bound of $(m-1)2^n + 2^{n-1}$ in the restricted case.
We will see later that both of these bounds are tight.

\section{Results}
\label{sec:main}
%



Let $\cA = (Q^\cA,\Sig^\cA,T^\cA,1',F^\cA)$ and $\cB = (Q^\cB,\Sig^\cB,T^\cB,1,F^\cB)$ be DFAs, with $Q^\cA = \{1',2',\dotsc,m'\}$ and $Q^\cB = \{1,2,\dotsc,n\}$ for positive integers $m$ and $n$. Let $\cC = (Q,\Sig,T,I,F)$ denote the concatenation DFA of $\cA$ and $\cB$ as defined in Section \ref{sec:def:aut}.

\br
Let $p',q' \in Q^\cA$, let $X,Y,Z \subseteq Q^\cB$, and let $w \in \Sig^*$.
Then:

\begin{quotation}
In $\cC$, if $(p',X)w = (q',Y)$, then $(p',X \cup Z)w = (q',Y \cup Zw)$.
\end{quotation}

Indeed, recall that the pair $(p',X)$ stands for the set $\{p'\} \cup X$. Thus $(\{p'\} \cup X)w = \{p'w\} \cup Xw = \{q'\} \cup Y$. It follows that $p'w = q'$ and $Xw = Y$. Hence $(\{p'\} \cup X \cup Z)w = \{p'w\} \cup Xw \cup Zw = \{q'\} \cup Y \cup Zw$, which in our pair notation is $(q',Y \cup Zw)$. 
We will readily use this basic fact in proofs. 
\er

Before stating our main result formally, we give some motivating exposition.
Fix a state $s' \in Q^\cA$ and a subset $B$ of $Q^\cB$.
The state $s'$ is called the \e{focus state} or simply \e{focus}; it is often taken to be the initial state $1'$ but in general can be any state.
The subset $B$ is called the \e{base}.
Fix a set $T$ with $B \subseteq T \subseteq Q^\cB$, called the \e{target}.
Our goal is to give sufficient conditions under which starting from $(s',B)$, we can reach $(s',S)$ for all sets $S$ with $B \subseteq S \subseteq T$. That is, we can reach any state of the concatenation DFA $\cC$ in which the first component is the focus and the second component lies between the base and the target.

The idea is to first assume we can reach $(s',B)$, the state consisting of the focus and the base.
Now, for $q \in Q$, define a \e{$q$-word} to be a word $w$ such that $(s',B)w = (s',B \cup q)$.
We can think of this as a word that ``adds'' the state $q$ to the base $B$.
Our next assumption is that we have a $q$-word for each state $q$ in the target $T$.
To reach a set $S$ with $B \subseteq S \subseteq T$, we will repeatedly use $q$-words to add each missing element of $S$ to the base $B$.

There is a problem with this idea, which we illustrate with an example.
Suppose $w_p$ is a $p$-word and $w_q$ is a $q$-word, and we want to reach $(s',B \cup \{p,q\})$.
Starting from $(s',B)$ we may apply $w_p$ to reach $(s',B \cup p)$.
But now if we apply $w_q$, we reach $(s',B \cup \{pw_q,q\})$.
There is no guarantee that we have $pw_q = p$, and in many cases we will not.
What we should really do is find a state $r$ such that $rw_q = p$, use an $r$-word to reach $(s',B \cup r)$, and then apply $w_q$ to reach $(s',B \cup \{p,q\})$.
But this idea only works if $p$ has a preimage under $w_q$, which may not be the case.

We resolve this by making a technical assumption, which ensures that preimages will always exist when we attempt constructions like the above.
First, define a \e{construction set} for the target $T$ to be a set of words consisting of exactly one $q$-word for each $q \in T$.
If $W$ is a construction set for $T$, we write $W[q]$ for the unique $q$-word in $W$.

We say a construction set is \e{complete} if there is a total order $\prec$ on the target $T$ such that for all $p,q \in T$ with $p \prec q$, the state $q$ has at least one preimage under the unique $p$-word $W[p]$, and at least one of these preimages lies in $T$. More formally, whenever $p \prec q$, the set $qW[p]^{-1} = \{s \in Q^\cB : sW[p] = q\}$ intersects $T$ non-trivially.
Our final assumption is that we have a complete construction set for $T$.

Note that the definition of a $q$-word depends not only on $q$, but also on $s'$ and $B$. Since a construction set for $T$ is a set of $q$-words, the definition of construction set also depends on $s'$ and $B$. For simplicity, we omit this dependence on $s'$ and $B$ from the notation for $q$-words and construction sets.

We summarize the definitions that have just been introduced:
\bi
\item
Fix a state  $s' \in Q^\cA$, called the \e{focus}, and a set $B \subseteq Q^\cB$ called the \e{base}.
\item
For $q \in Q^\cB$, a \e{$q$-word} is a word $w$ such that $(s',B)w = (s',B \cup q)$.
\item
Given a \e{target} set $T$ with $B \subseteq T \subseteq Q^\cB$, a \e{construction set} for $T$ is a set of words that contains exactly one $q$-word for each $q \in T$.
\item
The unique $q$-word in a construction set $W$ is denoted by $W[q]$.
\item
A construction set for $T$ is \e{complete} if there exists a total order $\prec$ on $T$ such that for all $p,q \in T$ with $p \prec q$, we have 
\[ qW[p]^{-1} \cap T = \{s \in Q^\cB : sW[p] = q\} \cap T \ne \emp. \]
\ei

Now, we state our main theorem, which gives the formal version of the construction described above.



\bt
\label{thm:reach-general}
Fix a state $s' \in Q^\cA$ and sets $B \subseteq T \subseteq Q^\cB$. 
If there is a complete construction set for $T$, 
then
all states of the form 
$(s',S)$ with $B \subseteq S \subseteq T$
are reachable from $(s',B)$ in $\cC$.
In particular, if $(s',B)$ itself is reachable, then
all states
$(s',S)$ with $B \subseteq S \subseteq T$
are reachable.
\et
\bpf
Note that if $B \subseteq S \subseteq T$, we can write $S = R \cup B$ with $R \cap B = \emp$ and $R \subseteq T$.
Thus it suffices to show that all states of the form $(s',R \cup B)$ with $R \cap B = \emp$ and $R \subseteq T$ are reachable from $(s',B)$.
We proceed by induction on $|R|$.
When $|R| = 0$, the only state of this form is $(s',B)$ itself.

Now suppose every state 
 $(s',R \cup B)$ with $R \cap B = \emp$, $R \subseteq T$ and $0 \le |R| < k$ is reachable from $(s',B)$.
We want to show this also holds for $|R| = k$.
Let $W$ be a complete construction set for $T$ and let $\prec$ be the corresponding total order on $T$.
Let $p$ be the minimal element of $R$ under $\prec$.
Let $w$ be $W[p]$, the unique $p$-word in $W$.
For all $q \in R \setminus p$, we have $p \prec q$ and thus $qw^{-1}$ contains an element of $T$ (since $W$ is complete).

Construct sets $X$ and $Y$ as follows: starting with $X = \emp$, for each $q \in R \setminus p$, choose an element of $qw^{-1} \cap T$ and add it to $X$.
Then set $Y = X \setminus B$.
Observe that $X$ is a subset of $T$ of size $|R \setminus p| = k-1$.
Hence $Y$ is a subset of $T$ of size at most $k-1$ such that $Y \cap B = \emp$.
It follows by the induction hypothesis that $(s',Y \cup B)$ is reachable from $(s',B)$.
But $Y \cup B = X \cup B$, so $(s',X \cup B)$ is reachable from $(s',B)$.
By the definition of $X$, we have $Xw = R \setminus p$.
Since $w$ is a $p$-word, we have $(s',B)w = (s',B \cup p)$, and thus
\[ (s',X \cup B)w = (s',Xw \cup B \cup p) = (s',(R \setminus p) \cup B \cup p) = (s',R \cup B). \]
Hence $(s',R \cup B)$ is reachable from $(s',B)$, as required.
\epf

The definition of completeness is somewhat complicated, which makes it difficult to use Theorem \ref{thm:reach-general}. Thus, we next prove some results giving useful sufficient conditions for a construction set to be complete.
Before stating our first such result, we introduce some notation.

Define $\Sig_0 = \Sig^\cA \cap \Sig^\cB$. We call $\Sig_0$ the \e{shared alphabet} of $\cA$ and $\cB$.
The following remark shows that when $\Sig^\cA \ne \Sig^\cB$, it is important to work exclusively with the shared alphabet when looking for complete construction sets. Of course, if $\Sig^\cA = \Sig^\cB$ then the shared alphabet is just the common alphabet of both automata, and there is nothing to worry about.

\br
A construction set for a non-empty target cannot be complete unless it is a subset of $\Sig_0^*$. To see this, suppose $W$ is a construction set and let $w \in W$. 
If $w$ contains a letter from $\Sig^\cA \setminus \Sig^\cB$, then $w$ is not a word over $\Sig^\cB$. Recall that if $w$ is not a word over $\Sig^\cB$, then $T^\cB_w$ is defined to be the empty relation. Thus the converse relation $(T^\cB_w)^{-1}$ is also empty, which means $qw^{-1}$ is empty for all $q$. It follows $W$ cannot be complete.
On the other hand, suppose $w$ contains a letter from $\Sig^\cB \setminus \Sig^\cA$. 
Then $(s',B)w = (\emp,Bw)$. Hence $w$ is not a $q$-word for any $q$, and so $w$ cannot be an element of a construction set, which is a contradiction. It follows that all words in a complete construction set are words over the shared alphabet $\Sig_0$.
\er


\bl
\label{lem:complete}
Fix $s' \in Q^\cA$ and sets $B \subseteq T \subseteq Q^\cB$.
Let $x_1,\dotsc,x_j$ be words over $\Sig_0$ that act as permutations on $T$, and
let $y$ be an arbitrary word over $\Sig_0$.
Choose $x_0 \in \{\eps,x_1,\dotsc,x_j\}$.
Define
\[ W = \{x_1,x_2,\dotsc,x_j\} \cup \{x_0y,x_0y^2,\dotsc,x_0y^k\}. \]
If $W$ is a construction set for $T$, then it is complete. 
\el
\bpf
For $1 \le i \le j$,
let $w_i = x_i$.
For $1 \le i \le k$,
let $w_{j+i} = x_0y^i$.
Let $\ell = j+k$.
Then we have $W = \{w_1,\dotsc,w_\ell\}$.
Let $q_i$ be the state in $T$ such that $(s',B)w_i = (s',B \cup q_i)$.
Define an order $\prec$ on $T$ so that $q_1 \prec q_2 \prec \dotsb \prec q_{\ell}$.
We claim this order makes $W$ complete.
Notice that $w_r = W[q_r]$, the unique $q_r$-word in $W$.
Thus we must show that whenever $q_r \prec q_s$, we have $q_sw_r^{-1} \cap T \ne \emp$.

Suppose $r < s$ and $r \le j$.
Then $w_r = x_r$ acts as a permutation on $T$.
Thus $q_sw_r^{-1} \cap T$ is non-empty, since $q_s \in T$.

Suppose $r < s$ and $r > j$.
Since $s-r > 0$, we can write $w_s = x_0y^{s-j} = x_0y^{s-r}y^{r-j} = w_{j+s-r}y^{r-j}$.
Thus $(s',B)w_s = (s',B \cup q_{j+s-r})y^{r-j} = (s',B \cup q_s)$.
There are two possibilities: $q_{j+s-r}y^{r-j} = q_s$, or $qy^{r-j} = q_s$ for some $q \in B$.

In either case, $q_s(y^{r-j})^{-1} \cap T$ is non-empty.
That is, there exists $q \in T$ such that $qy^{r-j} = q_s$.
Since $x_0$ acts as a permutation on $T$, there exists $p \in T$ such that $px_0 = q$.
Thus $px_0y^{r-j} = pw_r = q_s$.
It follows that $q_sw_r^{-1} \cap T$ is non-empty, as required.
\epf

Usually, we will use one of the following corollaries instead of Lemma \ref{lem:complete} itself.

\bc
\label{cor:complete}
Fix $s' \in Q^\cA$ and sets $B \subseteq T \subseteq Q^\cB$.
Let $x$ and $y$ be words over $\Sig_0$ such that $x$ acts as a permutation on $T$.
Suppose $W$ is one of the following sets:
\be
\item
$\{y,y^2,\dotsc,y^k\}$.
\item
$\{\eps,y,y^2,\dotsc,y^k\}$. 
\item
$\{x,xy,xy^2,\dotsc,xy^k\}$. 
\item
$\{\eps,x,xy,xy^2,\dotsc,xy^k\}$.
\ee
If $W$ is a construction set for $T$, then it is complete.
\ec
\bpf
All statements follow easily from Lemma \ref{lem:complete}:
\be
\item
Set $j = 0$.
\item
Set $j = 1$ and $x_0 = x_1 = \eps$.
\item
Set $j = 1$ and $x_0 = x_1 = x$.
\item
Set $j = 2$, $x_1 = \eps$ and $x_0 = x_2 = x$. \qedhere

\ee
\epf

\bc
\label{cor:complete-perm}
Fix $s' \in Q^\cA$ and sets $B \subseteq T \subseteq Q^\cB$.
Let $W \subseteq \Sig_0^*$ be a construction set for $T$.
\be
\item
If every word in $W$ acts as a permutation on $T$, then $W$ is complete.
\item
If there is a word $w \in W$ such that every word in $W \setminus w$ acts as a permutation on $T$, then $W$ is complete.
\ee
\ec
\bpf
Both statements follow easily from Lemma \ref{lem:complete}:
\be 
\item
Set $k = 0$ in Lemma \ref{lem:complete}.
\item
Set $k = 1$, $x_0 = \eps$ and $y = w$ in Lemma \ref{lem:complete}. \qedhere
\ee
\epf

In the special case where $W$ contains $\eps$, Corollary \ref{cor:complete-perm} admits the following generalization, which we found occasionally useful.
\bl
\label{lem:complete-perm}
Fix $s' \in Q^\cA$ and sets $B \subseteq T \subseteq Q^\cB$.
Let $W = \{\eps,w_1,\dotsc,w_k\}$ be a construction set for $T$, where $w_1,\dotsc,w_k$ are non-empty words over $\Sig_0$.
Suppose that for every word $w \in W$, there exists a set $S$ with $T \setminus B \subseteq S \subseteq T$ such that $w$ acts as a permutation on $S$.
Then $W$ is complete.
\el
\bpf
Write $B = \{q_1,\dotsc,q_j\}$.
Note that $\eps$ is a $q_i$-word for $1 \le i \le j$.
Thus by the definition of a construction set, $\eps$ is the unique $q_i$-word in $W$ for each $q_i \in B$, that is, $W[q_i] = \eps$ for $1 \le i \le j$.
In particular, each non-empty word in $W$ is a $q$-word for some $q \in T \setminus B$.
For $1 \le i \le k$, let $q_{j+i}$ be the state such that $(s',B)w_i = (s',B \cup q_{j+i})$.
Then $T = \{q_1,\dotsc,q_{j+k}\}$.
Note that $W[q_i] = \eps$ if $1 \le i \le j$, and $W[q_i] = w_{i-j}$ if $j+1 \le i \le j+k$.

Define an order $\prec$ on $T$ by $q_1 \prec q_2 \prec \dotsb \prec q_{j+k}$.
We claim this order makes $W$ complete.
Choose $q_r,q_s \in T$ with $q_r \prec q_s$; we want to show that $q_sW[q_r]^{-1} \cap T \ne \emp$.
Suppose $q_r \in B$. Then $W[q_r] = \eps$, and we have $q_s\eps^{-1} \cap T$ non-empty as required.
Now if $q_r \not\in B$, then since $q_r \prec q_s$ we also have $q_s \not\in B$.
In this case, $W[q_r] = w_{r-j}$, which acts as a permutation on some superset $S$ of $T \setminus B$. 
Since $q_s \in T \setminus B$, it follows that $q_s$ has a preimage under $w_{r-j}$, and furthermore this preimage lies in $T$, since $S$ is a subset of $T$.
Thus $q_sw_{r-j}^{-1} \cap T \ne \emp$ as required.
This proves that $W$ is complete.
\epf

Note that \e{all words} referred to in the above lemmas and corollaries are words over $\Sig_0$, the shared alphabet of $\cA$ and $\cB$. When working with automata that have different alphabets, it is important to only use words over the shared alphabet when trying to find a complete construction set.

The following ``master theorem'' summarizes all the results of this section. We have attempted to state this theorem in a form such that it can be cited without having to first define all the notions introduced in this section, such as $q$-words and construction sets and completeness. 
\bt
Let $\cA = (Q^\cA,\Sig^\cA,T^\cA,i^\cA,F^\cA)$ and $\cB = (Q^\cB,\Sig^\cB,T^\cB,i^\cB,F^\cB)$ be DFAs.
Let $\cC = (Q,\Sig,T,I,F)$ denote the concatenation DFA of $\cA$ and $\cB$, as defined in Section \ref{sec:def:aut}.
Let $\Sig_0 = \Sig^\cA \cap \Sig^\cB$.

Fix a state $s' \in Q^\cA$ and sets $B \subseteq T \subseteq Q^\cB$. 
Suppose that
for each $q \in T$, there exists a word $w_q \in \Sig_0^*$ such that $(s',B) \tr{w_q} (s',B \cup q)$ in $\cC$.
Let $W = \{w_q : q \in T\}$. 
Suppose that one of the following conditions holds:
\be
\item
There exist words $x,y \in \Sig_0^*$, where $x$ acts as a permutation on $T$, such that $W$ can be written in one of the following forms:
\bi
\item
$W = \{y,y^2,\dotsc,y^k\}$.
\item
$W = \{\eps,y,y^2,\dotsc,y^k\}$. 
\item
$W = \{x,xy,xy^2,\dotsc,xy^k\}$. 
\item
$W= \{\eps,x,xy,xy^2,\dotsc,xy^k\}$.
\ei
\item
Every word in $W$ acts as a permutation on $T$.
\item
There exists $w \in W$ such that every word in $W \setminus w$ acts as a permutation on $T$.
\item
$W$ contains $\eps$, and for every non-empty word $w \in W$, there exists a set $S$ such that $T \setminus B \subseteq S \subseteq T$ and $w$ acts as a permutation on $S$.
\item
There exists a total order $\prec$ on $T$ such that for all $p,q \in T$ with $p \prec q$, the set $qw_p^{-1} = \{ s \in Q^\cB : s \tr{w_p} q\}$ contains an element of $T$.
\ee
If one of the above conditions holds, then every state of the form $(s',X)$ with $B \subseteq X \subseteq T$ is reachable from $(s',B)$ in $\cC$.
\et

\section{Examples}
\label{sec:ex}
We now demonstrate our technique by applying it to various concatenation witnesses from the literature. 

\bt[Regular Language Witness. Brzozowski and Sinnamon, 2017~\cite{BrSi17}]
\label{thm:reg1}
Let $t \co Q^\cA \ra Q^\cA$ be a transformation such that $j't = 1'$.
Define $\cA$ and $\cB$ as follows:
\[
\begin{array}{lccc}
\ &a&b&\text{Final States}\\
\cA\co&(1',\dotsc,m')&t&\{m'\}\\
\cB\co&(1,\dotsc,n)&(2\ra 1)&\{n\}
\end{array}
\]
If $\gcd(j-1,n) = 1$, then $\cC$ has $(m-1)2^n+2^{n-1}$ reachable states.
In particular, transformations $t$ with $2't = 1'$ work for all $m$ and $n$.
\et
The authors of~\cite{BrSi17} proved this result with $t = (1',2')$, but we prove a slightly more general statement.
\bpf
The initial state of $\cC$ is $(1',\emp)$.
Set $x = a^m$ and $y = a^{j-1}b$.
We have
\[ (1',\emp) \tr{x} (1',2) \tr{y^k} (1',2+k(j-1)). \]
(Addition in the second component is performed modulo $n$.)
Since $j-1$ and $n$ are coprime, it follows from elementary number theory that $W=\{x,xy,\dotsc,xy^{n-1}\}$ is a construction set for $Q^\cB$ (with $s' = 1'$ and $B = \emp$).
By Corollary \ref{cor:complete}, it is complete.
Hence $(1',S)$ is reachable for all $S \subseteq Q^\cB$.
To reach $(q',S)$ for $q'$ non-final, first reach $(1',Sa^{-(q-1)})$ and then apply $a^{q-1}$.
To reach $(m',S \cup 1)$ for $S \subseteq Q^\cB \setminus 1$, first reach $((m-1)',Sa^{-1})$ and then apply $a$.
\epf
Note that the above theorem only gives conditions for $(m-1)2^n+2^{n-1}$ states to be \e{reachable}; it is not necessarily true that all the reachable states are pairwise distinguishable. For example, if $t$ is the constant transformation $(Q^\cA \ra 1')$ then $(p',Q^\cB)$ and $(q',Q^\cB)$ are indistinguishable. However, in~\cite{BrSi17} the authors take $t$ to be the transposition $(1',2')$ and find that all reachable states are pairwise distinguishable. 

In the remainder of our examples, all of the states we show are reachable will also be pairwise distinguishable. Since the focus of this paper is reachability, we refer to the original authors for distinguishability proofs in most cases. In cases where the original authors did not provide a distinguishability proof, we give a brief argument for completeness.


The next example involves two DFAs with different alphabets: we have $\Sig^\cA = \{a,b,c\}$ and $\Sig^\cB = \{a,b,d\}$. Our construction set will consist of words over the shared alphabet $\Sig_0 = \Sig^\cA \cap \Sig^\cB = \{a,b\}$.
\bt[Regular Language Witness. Brzozowski, 2016~\cite{Brz16}]
Define $\cA$ and $\cB$ as follows:
\[
\begin{array}{lccccc}
\ &a&b&c&d&\text{Final States}\\
\cA\co&(1',\dotsc,m')&(1',2')&(m' \ra 1')&\ &\{m'\}\\
\cB\co&(1,2)&(1,\dotsc,n)&\ &\id&\{n\}
\end{array}
\]
Then $\cC$ has $m2^n+2^{n-1}$ reachable and pairwise distinguishable states.
\et
\bpf
The initial state is $(1',\emp)$. 
If $n$ is odd, we have
\[ (1',\emp) \tr{a^m} (1',2) \tr{bb} (1',4) \tr{bb} \dotsb \tr{bb} (1',n-1), \]
\[ (1',n-1) \tr{bb} (1',1) \tr{bb} (1',3) \tr{bb} \dotsb \tr{bb} (1',n). \]
Thus $\{a^m,a^mbb,a^m(bb)^2,\dotsc,a^m(bb)^{n-1}\}$ is a construction set for $Q^\cB$ (with $s' = 1'$ and $B = \emp$). By Corollary \ref{cor:complete}, it is complete (taking $x = a^m$ and $y = bb$).

If $n$ is even, we have
\[ (1',\emp) \tr{a^m} (1',2) \tr{bb} (1',4) \tr{bb} \dotsb \tr{bb} (1',n), \]
\[ (1',n) \tr{ab} (1',1) \tr{bb} (1',3) \tr{bb} \dotsb \tr{bb} (1',n-1). \]
The words used to reach each state $(1',q)$ form a construction set for $Q^\cB$ (with $s' = 1$ and $B = \emp$).
We cannot use Corollary \ref{cor:complete} to show it is complete (since the appearance of $ab$ breaks the pattern), but notice that all the words in the construction set are words over $\{a,b\}$, and $a$ and $b$ both act as permutations on $Q^\cB$.
Thus all words in the construction set are permutations of $Q^\cB$, and so by Corollary \ref{cor:complete-perm} it is complete.

In either case, we have a complete construction set for $Q^\cB$ and so $(1',S)$ is reachable for all $S \subseteq Q^\cB$.
We can reach $(q',S)$ for $q' \ne m'$ and $(m',S \cup 1)$ by words in $a^*$, as in Theorem \ref{thm:reg1}.
This gives $(m-1)2^n+2^{n-1}$ reachable states.
Additionally, from $(q',S)$ we can reach $(\emp,S)$ by $d$, for an extra $2^n$ states.

For distinguishability of the reached states, see~\cite{Brz16}.
\epf
The main differences in reachability proofs between the different-alphabet case (unrestricted state complexity) and the same-alphabet case (restricted state complexity) are as follows:
\bi
\item
When looking for a complete construction set, we are restricted to using words over the shared alphabet $\Sig_0 = \Sig^\cA \cap \Sig^\cB$.
\item
Usually some additional states can be reached using letters in $\Sig^\cA \setminus \Sig^\cB$ or $\Sig^\cB \setminus \Sig^\cA$, e.g., the states of the form $(\emp,S)$ in the previous example.
\ei
As these differences are not too significant, we will stick to the same-alphabet case for the remainder of our examples.

\bt[Regular Language Witness. Brzozowski, 2013~\cite{Brz13}]
\label{thm:reg2}
Define $\cA$ and $\cB$ as follows:
\[
\begin{array}{lcccc}
\ &a&b&c&\text{Final States}\\
\cA\co&(1',\dotsc,m')&(1',2')&(m' \ra 1')&\{m'\}\\
\cB\co&(1,\dotsc,n)&(1,2)&(n \ra 1)&\{n\}
\end{array}
\]
Then $\cC$ has $(m-1)2^n+2^{n-1}$ reachable and pairwise distinguishable states.
\et
\bpf
The initial state of $\cC$ is $(1',\emp)$. 
For $0 \le k \le n-2$ we have
\[ (1',\emp) \tr{a^m} (1',2) \tr{(ab)^k} (1',2+k). \]
Also, $(1',n) \tr{c} (1',1)$.
Thus 
$\{a^m,a^mab,a^m(ab)^2,\dotsc,a^m(ab)^{n-2},a^m(ab)^{n-2}c\}$
is a construction set for $Q^\cB$ (with $s'=1'$ and $B = \emp$). 

This construction set does not quite have the right form to apply Corollary \ref{cor:complete}, due to the last word $a^m(ab)^{n-2}c$.
However, notice that all words in $W$ except for $a^m(ab)^{n-2}c$ are in fact permutations of $Q^\cB$, so Corollary \ref{cor:complete-perm} shows that $W$ is complete.
Hence all states $(1',S)$ with $S \subseteq Q^\cB$ are reachable.
We can reach $(q',S)$ for $q' \ne m'$ and $(m',S \cup 1)$ by words in $a^*$, as in Theorem \ref{thm:reg1}.

For distinguishability of the reached states, see~\cite{Brz13}.
\epf

\bt[Regular Language Witness. Yu, Zhuang and Salomaa, 1994~\cite{YZS94}]
Define $\cA$ and $\cB$ as follows:
\[
\begin{array}{lcccc}
\ &a&b&c&\text{Final States}\\
\cA\co&(1',\dotsc,m')&(Q^\cA \ra 1')&\id&\{m'\}\\
\cB\co&\id&(1,\dotsc,n)&(Q^\cB \ra 2)&\{n\}
\end{array}
\]
Then $\cC$ has $(m-1)2^{n}+2^{n-1}$ reachable and pairwise distinguishable states.
\et
\bpf
The initial state of $\cC$ is $(1',\emp)$.
For $k \le n-2$ we have
$(1',\emp) \tr{a^m} (1',2) \tr{b^k} (1',2+k)$,
and $(1',n) \tr{b} (1',1)$.
It follows that $\{a^m,a^mb,\dotsc,ab^{n-1}\}$ is a construction set for $Q^\cB$ (with $s' = 1'$ and $B = \emp$).
By Corollary \ref{cor:complete}, it is complete (taking $x = a^m$ and $y = b$).
Hence all states $(1',S)$ with $S \subseteq Q^\cB$ are reachable.
We can reach $(q',S)$ for $q' \ne m'$ and $(m',S \cup 1)$ by words in $a^*$.


Let $(p',S)$ and $(q',T)$ be distinct states of $\cC$.
If $S \ne T$, let $r$ be a state in the symmetric difference of $S$ and $T$. Then $b^{n-r}$ distinguishes the states. If $S = T$ and $p' < q'$, then $ca^{m-q}b^{n-2}$ distinguishes the states.
\epf

\bt[Regular Language Witness. Maslov, 1970~\cite{Mas70}]
Define $\cA$ and $\cB$ as follows:
\[
\begin{array}{lcccc}
\ &a&b&\text{Final States}\\
\cA\co&(1',\dotsc,m')&\id&\{m'\}\\
\cB\co&(n-1,n)&({}_1^{n-1} q \ra q+1)&\{n\}
\end{array}
\]
Then $\cC$ has $(m-1)2^{n}+2^{n-1}$ reachable and pairwise distinguishable states.
\et
\bpf
The initial state is $(1',\emp)$. We have
\[ (1',\emp) \tr{a^m} (1',1) \tr{b^k} (1',1+k). \]
Thus $\{a^m,a^mb,a^mb^2,\dotsc,a^mb^{n-1}\}$ is a construction set for $Q^\cB$ (with $s' = 1'$ and $B = \emp$).
By Corollary \ref{cor:complete}, it is complete. Hence $(1',S)$ is reachable for all $S \subseteq Q^\cB$.
We can reach $(q',S)$ for $q' \ne m'$ and $(m',S \cup 1)$ by words in $a^*$, as in Theorem \ref{thm:reg1}.

Let $(p',S)$ and $(q',T)$ be distinct states of $\cC$.
If $S \ne T$, let $r$ be a state in the symmetric difference of $S$ and $T$. Then $b^{n-r}$ distinguishes the states. If $S = T$ and $p' < q'$, by $b^n$ we reach $(p',n)$ and $(q',n)$. Then by $a^{m-q}$ we reach $((p+m-q)',na^{m-q})$ and $(m',na^{m-q} \cup 1)$. These states differ in their second component, so they are distinguishable.
\epf

\bt[Star-Free Witness. Brzozowski and Liu, 2012~\cite{BrLi12}]
Define $\cA$ and $\cB$ as follows:
\[
\begin{array}{lcccc}
\ &a&b&c&d\\
\cA\co&({}_1^{m-1} q' \ra (q+1)')&({}_2^m q' \ra (q-1)')&\id&(Q^\cA \ra m')\\
\cB\co&({}_2^{n-1} q \ra q+1)&\id&({}_1^{n-1} q \ra q+1)&({}_2^n q \ra q-1)\\
\end{array}
\]
and let $F^\cA = \{m'\}$ and $F^\cB = \{n-1\}$.
Then $\cC$ has $(m-1)2^{n}+2^{n-1}$ reachable and pairwise distinguishable states.
\et
\bpf
The initial state is $(1',\emp)$.
We have
\[ (1',\emp) \tr{a^{m}} (m',1) \tr{c^k} (m',\{1,1+k\}). \]
Hence $\{\eps,c,c^2,\dotsc,c^{n-1}\}$ is a construction set for $Q^\cB$ (with $s' = m'$ and $B = \{1\}$).
By Corollary \ref{cor:complete}, it is complete.
Thus $(m',S \cup 1)$ is reachable for all $S \subseteq Q^\cB$.

To reach $(q',S)$ for non-final $q' \in Q^\cA$ and $S \subseteq Q^\cB$, proceed as follows.
If $1 \in S$, first reach $(m',S \cup 1)$ then apply $b^{m-q}$.
If $1 \not\in S$, let $i$ be the smallest element of $S$.
Set $T = \{q-(i-1) : q \in S \setminus i\}$ and reach $(m',T \cup 1)$.
Then $(m',T \cup 1) \tr{b^{m-q}} (q',T \cup 1) \tr{c^{i-1}} (q',(S\setminus i) \cup i) = (q',S)$.

For distinguishability of the reached states, see~\cite{BrLi12}.
\epf

\bt[Non-Returning Witness. Brzozowski and Davies, 2017~\cite{BrDa17}]
Define $\cA$ and $\cB$ as follows:
\[
\begin{array}{lccc}
\ &a&b&\text{Final States}\\
\cA\co&(2',\dotsc,m')(1'\ra 2')&(2',3')(1' \ra 3')&\{m'\}\\
\cB\co&(2,\dotsc,n)(1 \ra 2)&(3,\dotsc,n)(2 \ra 3)(1 \ra 2)&\{n\}
\end{array}
\]
Then $\cC$ has $(m-1)2^{n-1}+1$ reachable and pairwise distinguishable states.
\et
\bpf
The initial state is $(1',\emp)$.
Let $x = a^{m-1}$ and $y = ab$.
If $n$ is even,
\[ 
(1',\emp) \tr{a} (2',\emp) \tr{x} (2',2) \tr{y} (2',4) \tr{y} (2',6) \tr{y} \dotsb \tr{y} (2',n), \]
\[ (2',n) \tr{y} (2',3) \tr{y} (2',5) \tr{y} \dotsb \tr{y} (2',n-1). \]
If $n$ is odd,
\[ 
(1',\emp) \tr{a} (2',\emp) \tr{x} (2',2) \tr{y} (2',4) \tr{y} (2',6) \tr{y} \dotsb \tr{y} (2',n-1), \]
\[ (2',n-1) \tr{y} (2',3) \tr{y} (2',5) \tr{y} \dotsb \tr{y} (2',n). \]
In both cases, Corollary \ref{cor:complete} implies that $\{x,xy,\dotsc,xy^{n-2}\}$ is a complete construction set for $Q^\cB \setminus 1$ (with $s'=2'$ and $B = \emp$).
It follows that $(2',S)$ is reachable for all $S \subseteq Q^\cB \setminus 1$. This gives $2^{n-1}$ reachable states.

To reach $(q',S)$ for non-final $q' \in Q^\cA \setminus 1$ and $S \subseteq Q^\cB \setminus 1$, note that $a$ acts as a permutation on $Q^\cB \setminus 1$, and so there exists $T \subseteq Q^\cB \setminus 1$ such that $Ta^{q-2} = S$. Thus we can first reach $(2',T)$ and then apply $a^{q-2}$.
To reach $(m',S \cup 1)$ for $S \subseteq Q^\cB \setminus 1$, reach $((m-1)',T)$ where $Ta = S$ and apply $a$.
Counting the initial state $(1,\emp)$, we get $(m-1)2^{n-1} + 1$ reachable states.

For distinguishability of the reached states, see~\cite{BrDa17}.
\epf

\bt[Non-Returning Witness. Eom, Han and Jir\'askov\'a, 2016~\cite{EHJ16}]
Define $\cA$ and $\cB$ as follows:
\[
\begin{array}{lccc}
\ &a&b&c\\
\cA\co&(2',\dotsc,m')(1'\ra 2')&(1' \ra 2')&(1' \ra 2')\\
\cB\co&(1 \ra 2)&(2,\dotsc,n)(1 \ra 2)&({}_3^{n-1} q \ra q+1)(1 \ra 2)(n \ra 2)\\
\end{array}
\]
and let $F^\cA = \{m'\}$ and $F^\cB = \{n\}$.
Then $\cC$ has $(m-1)2^{n-1}+1$ reachable and pairwise distinguishable states.
\et
\bpf
The initial state is $(1',\emp)$.
We have
\[ (1',\emp) \tr{a} (2',\emp) \tr{a^{m-1}} (2',2) \tr{b^k} (2',2+k). \]
Hence by Corollary \ref{cor:complete}, $\{a^{m-1},a^{m-1}b,\dotsc,a^{m-1}b^{n-2}\}$ is a complete construction set for $Q^\cB \setminus 1$ (with $s'=2'$ and $B = \emp$).
It follows that $(2',S)$ is reachable for all $S \subseteq Q^\cB \setminus 1$.
To reach $(q',S)$ for $q'$ non-final, reach $(2',S)$ and apply $a^{q-2}$.
For $(m',S \cup 1)$, reach $((m-1)',S)$ and apply $a$.

For distinguishability of the reached states, see~\cite{EHJ16}.
\epf


\bt[Prefix-Closed Witness. Brzozowski, Jir\'askov\'a and Zou, 2014~\cite{BJZ14}]
Define $\cA$ and $\cB$ as follows:
\[
\begin{array}{lcccc}
\ &a&b&c&\text{Final States}\\
\cA\co&\id&\id&({}_1^{m-1} q' \ra (q+1)')&\{1',\dotsc,(m-1)'\}\\
\cB\co&(1,\dotsc,n-1)&({}_2^{n-1} q \ra q+1)&\id &\{1,\dotsc,n-1\}
\end{array}
\]
Then $\cC$ has $(m+1)2^{n-2}$ reachable and pairwise distinguishable states.
\et
\bpf
The initial state is $(1',1)$. 
For $k \le n-2$ we have $(1',1) \tr{a^k} (1',\{1,1+k\})$. 
Thus by Corollary \ref{cor:complete} the set $\{\eps,a,a^2,\dotsc,a^{n-2}\}$ is a complete construction set for $Q^\cB \setminus n$, with $s' = 1'$ and $B = \{1\}$.
Hence $(1',S \cup 1)$ is reachable for each $S \subseteq Q^\cB \setminus n$.
From $(1',S \cup 1)$ with $S \subseteq Q^\cB \setminus n$, we reach $(q',S \cup 1)$ for $2 \le q \le m$ by $c^{q-1}$.
This gives $m2^{n-2}$ reachable states.

To reach $(m',S)$ with $S \subseteq Q^\cB \setminus n$, set $S$ non-empty, and $1 \not\in S$, let $p$ be the smallest element of $S$. Let $T = Sa^{-(p-1)}$; then $1 \in T$ since $1a^{p-1} = p$. Reach $(m',T)$ and apply $a^{p-1}$ to reach $(m',S)$. There are $2^{n-2}-1$ non-empty sets that exclude $1$ and $n$, and we can reach an additional state $(m',n)$ from $(m',n-1)$ by $b$. This gives another $2^{n-2}$ reachable states, for a total of $(m+1)2^{n-2}$ states.

For distinguishability of the reached states, see~\cite{BJZ14}. 
\epf

\bt[Suffix-Free Witness. Brzozowski and Sinnamon, 2017~\cite{BrSi17a}]
Define $\cA$ and $\cB$ as follows:
\[
\begin{array}{lccc}
\ &a&b&c\\
\cA\co&(1'\ra m')(2',\dotsc,(m-1)')&(1'\ra m')(2',3')&(2',m')(1' \ra 2')\\
\cB\co&(1 \ra n)(2,3)&(2,n)(1 \ra 2)&(1 \ra n)(2,\dotsc,n-1)\\
\end{array}
\]
and let $F^\cA = \{(m-1)'\}$ and $F^\cB = \{n-1\}$.
Then $\cC$ has $(m-1)2^{n-2}+1$ reachable and pairwise distinguishable states.
\et
\bpf
The initial state is $(1',\emp)$. 
We have
\[ (1',\emp) \tr{c} (2',\emp) \tr{a^{m-3}} ((m-1)',1) \tr{c} ((m-1)',\{1,n\}). \]
Then for $k \le n-3$ we have
\[ ((m-1)',\{1,n\}) \tr{bb} ((m-1)',\{1,2,n\}) \tr{c^k} ((m-1)',\{1,2+k,n\}). \]
Thus $W = \{\eps,bb,bbc,bbc^2,\dotsc,bbc^{n-3}\}$ is a construction set for $Q^\cB$, with $s' = (m-1)'$ and $B = \{1,n\}$.
In fact, $W$ is complete by Lemma \ref{lem:complete-perm} since $b$ and $c$ act as permutations on $Q^\cB \setminus 1$.

It follows that $((m-1)',S \cup \{1,n\})$ is reachable for all $S \subseteq Q^\cB$.
To reach $(q',S \cup n)$ for $2 \le q \le m-2$ and $1 \not\in S$, note that $a$ acts as a permutation on $Q^\cB \setminus 1$. Thus we first reach $((m-1)',Sa^{-(q-1)} \cup \{1,n\})$ then apply $a^{q-1}$.
To reach $(m',S \cup n)$ with $1 \not\in S$, first reach $(2',Sc^{-1} \cup n)$ then apply $c$.
Since there are $2^{n-2}$ subsets of $Q^\cB \setminus \{1,n\}$, this gives $(m-1)2^{n-2}$ reachable states. Adding one for the initial state $(1',\emp)$ gives $(m-1)2^{n-2}+1$.

For distinguishability of the reached states, see~\cite{BrSi17a}. 
Note that the authors of~\cite{BrSi17a} use a different concatenation DFA from our $\cC$: they first delete the sink states $m'$ from $\cA$ and $n$ from $\cB$, and then form the concatenation of these modified DFAs. However, the same words used for distinguishing states in~\cite{BrSi17a} can be used to distinguish states of $\cC$. 
\epf

\bt[Suffix-Free Witness. Han and Salomaa, 2009~\cite{HaSa09}]
Define $\cA$ and $\cB$ as follows:
\[
\begin{array}{lcc}
\ &\cA\co&\cB\co\\
a&(2',\dotsc,(m-1)')(1'\ra m')&(1 \ra n)\\
b&(1'\ra m')&(2,\dotsc,n-1)(1 \ra n)\\
c&((Q^\cA \setminus 1') \ra m')(1' \ra 2')&(1 \ra n)\\
d&((Q^\cA \setminus 2') \ra m')&(1 \ra 2)\\
\end{array}
\]
and let $F^\cA = \{2'\}$ and $F^\cB = \{2\}$.
Then $\cC$ has $(m-1)2^{n-2}+1$ reachable and pairwise distinguishable states.
\et
\bpf
The initial state is $(1',\emp)$.
For $k \le n-3$ we have
\[ (1',\emp) \tr{cb} (2',\{1,n\}) \tr{d} (2',\{1,2,n\}) \tr{b^k} (2',\{1,2+k,n\}). \]
Thus $W = \{\eps,d,db,\dotsc,db^{n-3}\}$ is a construction set for $Q^\cB$, with $s' = 2'$ and $B = \{1,n\}$.
By Lemma \ref{lem:complete-perm}, $W$ is complete, since $d$ and $b$ act as permutations on $Q^\cB \setminus \{1,n\}$.


There are $2^{n-2}$ states of the form $(2',S \cup \{1,n\})$ with $S \subseteq Q^\cB$ and $S \cap \{1,n\} = \emp$. For each of these states, we reach $(q',S \cup n)$ for $3 \le q \le m-1$ by $a^{q-2}$, and $(m',S \cup n)$ by $c$. Adding in the initial state $(1',\emp)$ gives a total of $(m-1)2^{n-2}+1$ reachable states. 

For distinguishability of the reached states, see~\cite{HaSa09}. 
Note that the authors of~\cite{HaSa09} work with a reduced concatenation DFA obtained by identifying, for each $q'$ and $S$, the indistiguishable states $(q',S)$ and $(q',S \cup n)$. Thus, for example, they write that $(q',\emp)$ is reachable for $3 \le q \le m-1$; these states are not reachable in our DFA $\cC$, but states $(q',n)$ for $3 \le q \le m-1$ are reachable.
\epf

\bt[Right Ideal Witness. Brzozowski and Sinnamon, 2017~\cite{BrSi17}]
Define $\cA$ and $\cB$ as follows:
\[
\begin{array}{lcccc}
\ &a&b&c&\text{Final States}\\
\cA\co&(1',\dotsc,(m-1)')&(2'\ra1')&({}_1^{m-1} q' \ra (q+1)')&\{m'\}\\
\cB\co&(1,\dotsc,n-1)&(2 \ra 1)&({}_1^{n-1} q \ra q+1) &\{n\}
\end{array}
\]
Then $\cC$ has $m+2^{n-2}$ reachable and pairwise distinguishable states.
\et
\bpf
The initial state is $(1',\emp)$.
Note that $(1',\emp) \tr{a^{q-1}} (q',\emp)$ for $1 \le q \le m-1$, so these $m-1$ states are reachable.
For $0 \le k \le n-3$ we have 
\[ ((m-1)',\emp) \tr{c} (m',1) \tr{a} (m',\{1,2\}) \tr{(ab)^k} (m',\{1,2+k\}).\]
Hence $\{\eps,a,aab,a(ab)^2,\dotsc,a(ab)^{n-3}\}$ is a construction set for $Q^\cB \setminus n$, with $s' = m'$ and $B = \{1\}$.
By Corollary \ref{cor:complete}, it is complete.
Hence $(m',S \cup 1)$ is reachable for all $S \subseteq Q^\cB \setminus n$.

We have reached $(m-1) + 2^{n-2}$ states so far.
Additionally, we have
$(m',\{1,n-1\}) \tr{cb} (m',\{1,n\})$, giving $m+ 2^{n-2}$.



For distinguishability of the reached states, see~\cite{BrSi17}.
\epf

\bt[Right Ideal Witness. Brzozowski, Davies and Liu, 2016~\cite{BDL16}]
Define $\cA$ and $\cB$ as follows:
\[
\begin{array}{lcccc}
\ &a&b&c&\text{Final States}\\
\cA\co&(1',\dotsc,(m-1)')&(2',\dotsc,(m-1)')&((m-1)' \ra m')&\{m'\}\\
\cB\co&(1,\dotsc,n-1)&(2,\dotsc,n-1)&(n-1 \ra n)&\{n\}
\end{array}
\]
Then $\cC$ has $m+2^{n-2}$ reachable and pairwise distinguishable states.
\et
\bpf
The initial state is $(1',\emp)$.
Note that $(1',\emp) \tr{a^{q-1}} (q',\emp)$ for $1 \le q \le m-1$, so these $m-1$ states are reachable.
For $0 \le k \le n-3$ we have 
\[ ((m-1)',\emp) \tr{c} (m',1) \tr{a} (m',\{1,2\}) \tr{b^k} (m',\{1,2+k\}). \]
Hence $\{\eps,a,ab,ab^2,\dotsc,ab^{n-3}\}$ is a construction set for $Q^\cB \setminus n$, with $s' = m'$ and $B = \{1\}$. By Corollary \ref{cor:complete}, it is complete. 
Hence $(m',S \cup 1)$ is reachable for all $S \subseteq Q^\cB \setminus n$.

We have reached $(m-1) + 2^{n-2}$ states so far.
Additionally, we have
$(m',\{1,n-1\}) \tr{c} (m',\{1,n\})$, giving $m+ 2^{n-2}$.

For distinguishability of the reached states, see~\cite{BDL16}.
\epf


\bt[Right Ideal Witness. Brzozowski, Jir\'askov\'a and Li, 2013~\cite{BJL13}]
Define $\cA$ and $\cB$ as follows:
\[
\begin{array}{lccc}
\ &a&b&\text{Final States}\\
\cA\co&({}_1^{m-1} q' \ra (q+1)')&({}_1^{m-1} q' \ra (q+1)')&\{m'\}\\
\cB\co&(1,\dotsc,n-1)&({}_2^{n-1} q \ra q+1)&\{n\}
\end{array}
\]
Then $\cC$ has $m+2^{n-2}$ reachable and distinguishable states.
\et
\bpf
The initial state is $(1',\emp)$.
Note that $(1',\emp) \tr{a^{q-1}} (q',\emp)$ for $1 \le q \le m-1$, so these $m-1$ states are reachable.
For $0 \le k \le n-3$ we have 
\[ ((m-1)',\emp) \tr{a} (m',1) \tr{a} (m',\{1,2\}) \tr{b^k} (m',\{1,2+k\}).\]
Hence $\{\eps,a,ab,ab^2,\dotsc,ab^{n-3}\}$ is a construction set for $Q^\cB \setminus n$, with $s' = m'$ and $B = \{1\}$. By Corollary \ref{cor:complete}, it is complete. 
Hence $(m',S \cup 1)$ is reachable for all $S \subseteq Q^\cB \setminus n$.

We have reached $(m-1) + 2^{n-2}$ states so far.
Additionally, we have
$(m',\{1,n-1\}) \tr{b} (m',\{1,n\})$, giving $m+ 2^{n-2}$.

For distinguishability of the reached states, see~\cite{BJL13}. Note that the authors of~\cite{BJL13} use a different concatenation DFA, constructed by removing state $m'$ from $\cA$ and then forming the concatenation in the usual way. However, the same words used in~\cite{BJL13} can be used to distinguish states in $\cC$.
\epf

We now give two examples where our method of proof does not seem applicable or helpful. When attempting concatenation state complexity proofs, it seems best to consider both traditional techniques and the technique we present in this paper, switching between the two options if one does not yield an easy argument.

\bx[Prefix-Closed Witness. Brzozowski and Sinnamon, 2017~\cite{BrSi17}]
\label{ex:bad}
Our technique does not seem to work well with the following witness languages.
Define $\cA$ and $\cB$ as follows:
\[
\begin{array}{lcc}
\ &\cA\co&\cB\co\\
a&(1',\dotsc,(m-1)')&(1,\dotsc,n-1)\\
b&(1',2')&(2 \ra 1)\\
c&(2' \ra 1')&({}_1^{n-1} q \ra q-1)\\
d&({}_1^{m-1} q' \ra (q-1)')&(1,2)\\
\end{array}
\]
and let $F^\cA = \{1',\dotsc,(m-1)'\}$ and $F^\cB = \{1,\dotsc,n-1\}$.

The inductive proof given by the authors of~\cite{BrSi17} has a different structure from the type of argument captured by Theorem \ref{thm:reach-general}. 
To reach a state $(q',S)$, in Theorem \ref{thm:reach-general} we start from some state $(q',B)$ and apply a word that fixes the first component $q'$. 
In~\cite{BrSi17} the authors instead start from a state $(p',B)$ and apply a word $w$ such that $p'w = q'$. The proof in~\cite{BrSi17} is short and clean, whereas a proof in the style of Theorem \ref{thm:reach-general} seems to require complicated arguments.
It is possible that Theorem \ref{thm:reach-general} could be generalized to cover arguments of the form used in~\cite{BrSi17}, but we have not found such a generalization.
\ex

\bx[Finite Binary Witness. C\^{a}mpeanu, Culik, Salomaa and Yu, 2001~\cite{CCSY01}]
\label{ex:finite}
Our technique does not apply to the following witness languages.
Define $\cA$ and $\cB$ as follows:
\[
\begin{array}{lcccc}
\ &a&b&\text{Final States}\\
\cA\co&({}_1^{m-1} q' \ra (q+1)')&({}_1^{m-1} q' \ra (q+1)')&\{1',\dotsc,(m-1)'\}\\
\cB\co&({}_2^{n-1} q \ra q+1)(1 \ra n)&({}_1^{n-1} q \ra q+1) &\{n-1\}
\end{array}
\]
Additionally, assume that $m+1 \ge n > 2$.
Then $\cC$ has $(m-n+3)2^{n-2}-1$ reachable and pairwise distinguishable states. This is the maximum for finite languages over a binary alphabet when $m+1 \ge n > 2$. 


Let us consider why Theorem \ref{thm:reach-general} cannot be used here. The point of Theorem \ref{thm:reach-general} is to build up states $(s',S)$ by starting from $(s',B)$ and using words that fix the focus state $s'$. But in this witness, no state of $\cA$ is fixed by any word except for the non-final sink state $m'$. So to use Theorem \ref{thm:reach-general}, the focus state must be $m'$. But from a state of the form $(m',S)$, we can only reach sets $(m',T)$ with $|T| \le |S|$, since $m'$ is a non-final sink state. So there is no way to start from some base state $(m',B)$ and build up larger sets, which is the strategy of Theorem \ref{thm:reach-general}.
\ex

%



\section{Conclusions}
\label{sec:con}
We have introduced a new technique for demonstrating the reachability of states in DFAs for the concatenation of two regular languages, and provided evidence that this technique is useful in a wide variety of cases. 
However, we found two cases (Examples \ref{ex:bad} and \ref{ex:finite}) where our technique does not seem applicable. Example \ref{ex:bad} in particular suggests that Theorem \ref{thm:reach-general} may admit a generalization that covers more types of inductive proofs. We leave this as an open problem.

\subsection*{Acknowledgements} 
I thank Jason Bell and Janusz Brzozowski for proofreading and helpful comments.  
This work was supported by the Natural Sciences and Engineering
Research Council of Canada under grant No.\ OGP0000871.


\bibliographystyle{abbrv}
\bibliography{concatenation}
\end{document}